\documentclass[3p, times, twocolumn]{elsarticle}

\usepackage{amssymb}

\usepackage{lineno}

\usepackage[figuresright]{rotating}

\begin{document}

\begin{frontmatter}




\title{Research and calibration of Acoustic Sensors in ice within the SPATS (South Pole Acoustic Test Setup) project}


\author[label1]{Thomas Meures}
\author[label1]{Larissa Paul}
\author[label2]{Mathieu Ribordy}
\author[label3]{for the IceCube collaboration}

\address[label1]{III Physikalisches Institut, RWTH Aachen University, D-52064	Aachen, Germany}
\address[label2]{Laboratory for High Energy Physics, $\acute{E}$cole Polytechnique F$\acute{e}$d$\acute{e}$rale, CH-1015 Lausanne, Switzerland}
\address[label3]{http://icecube.wisc.edu}

\begin{abstract}
We present development work aiming towards a large scale ice-based hybrid detector including acoustic sensors for the detection of neutrinos in the GZK range. A facility for characterization and calibration of acoustic sensors in clear (bubble-free) ice has been developed and the first measurements done at this facility are presented. Further, a resonant sensor intended primarily for characterization of the ambient noise in the ice at the South Pole has been developed and some data from its performance are given.
\end{abstract}

\begin{keyword}
Acoustic neutrino detection \sep SPATS


\end{keyword}

\end{frontmatter}

\section{Introduction}
An important supplement to the in situ test at South Pole with the South Pole Acoustic Test Setup (SPATS) \cite{Boeser:2006fx} are the laboratory activities at all contributing institutions. We describe here first the facility for measurements on sensors and transmitters frozen into a 3$\,$m$^3$ high quality ice-block. The facility will also be used for measurements with thermo-acoustically generated pulses, but this will not be described below. We further describe a resonant sensor that has been developed primarily for determination of the ambient noise level in the ice at the South Pole. It may also be a possible sensor in a complementary part of a radio array intended for detection of GZK neutrinos at the South Pole.
\section{The Aachen Acoustic Laboratory}
\label{AAL}
\subsection{AAL facility}
\label{AALf}
In the Aachen Acoustic Laboratory (AAL), an IceTop tank has been set up inside a cooling container for calibration of acoustic sensors in ice. In the cylindrical tank, with a diameter of 190$\,$cm and a used height of 85$\,$cm, bubble-free ice can be produced with the help of a freeze control unit (FCU) \cite{IceTop}. The FCU is equipped with a vacuum pump connected to a semi-permeable membrane, in order to degas the water. The water in the tank freezes starting from the top. To take out the surplus water under the ice, an overflow reservoir is used.  The total freezing of the 3$\,$m$^3$ of water takes around 70 days.\\
Inside the tank, 18 piezo-based sender/sensor-pairs are mounted on a dodecagonal aluminum frame embedded in the ice volume. Each of the three levels holds six sender/sensor pairs in a hexagon structure. The levels are located at 5$\,$cm, 25$\,$cm, and 45$\,$cm below the ice surface. The third level is situated 40$\,$cm above the bottom of the tank. The used senders and sensors are simple Piezoelectric Transducer (PZT) discs in an aluminum housing produced by Murata \cite{Murata}. The sensors are equipped with  a pre-amplifier with a gain of 100 and a differential output. They are calibrated absolutely by means of the reciprocity method  and serve as reference for calibration of other acoustic devices. The accuracy of the positioning system was checked during the freezing using a $\chi^2$ minimization of the arrival times of acoustic signals produced by the senders \cite{Simon}. The  positions were reconstructed with an accuracy of 0.5$\,$cm.
\subsection{Reciprocity calibration in the AAL}
\label{recalibration}
The sensors, used in the AAL setup as calibration references, are themselves calibrated by means of the reciprocity method \cite{Urick}. This method is based on the reciprocity principle, developed for systems such as electro-acoustic converters. It leads to the reciprocity equation:
\begin{eqnarray}
 M = J S.
\end{eqnarray}

\begin{figure}[ht]
	\begin{center}
	\includegraphics[width=4.5cm]{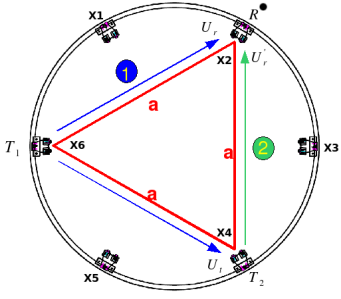}
	\caption{ A schematic setup for the calibration of one sensor $R$ with two transmitters $T_1$ and $T_2$.}	
	\label{fig_calibration_ring}
	\end{center}
\end{figure}

Here, $M = U/p$ is the sensitivity of the converter. It is the ratio of a response voltage $U$ at the sensor and the incident pressure p, causing the sensor response. The emissivity $S=p_0/I$ is the ratio of an emitted pressure at a distance of one meter and the applied current to the converter \cite{Lerch}. The reciprocity coefficient $J$ has to be determined individually for each sensor type. We calculated $J$ by using the baffled piston model \cite{Lerch}. The coefficient is:
\begin{eqnarray}
 J = \frac{2 \pi r}{\omega \rho_0} \cdot \frac{1}{\lambda_T(\alpha, \omega)}
\end{eqnarray}
Here $r$ is the distance between the sender and the sensor, $\omega$ the angular frequency of a signal, $\rho_0$ the density of the medium and $\lambda_T(\alpha, \omega)$ the angular dependence of the converter sensitivity and emissivity for the angle $\alpha$ and angular frequency $\omega$.\\
The sensors are mounted in the corners of a hexagon, such that an equilateral triangle is formed by three sensors. For the reciprocity calibration of a sensor R (Fig. \ref{fig_calibration_ring}) a transmitter T$_1$ on one corner of the triangle transmits to the sensor and the second sensor/transmitter T$_2$ with a broad band signal. Being at the same distance both sensors receive the same sound field. The output voltages $U_R$ and $U_T$ are recorded. Subsequently, a current $I$ is applied to the reference transmitter T$_2$ which causes an acoustic signal to be sent, recorded by R and resulting in an output voltage $U_R'$. The sensitivity can then be determined by

\begin{eqnarray}
 M_R = \frac{1}{\lambda_R(\alpha, \omega)} \cdot \sqrt{\frac{U_R U_R' 2 \pi r}{U_T I_T \omega \rho_0}}.
\end{eqnarray}

Here $\lambda_R$ the angular dependence of the sensor . To calculate the current I, applied to the reference transmitter, its impedance has to be measured.  For an electro-acoustic converter an electric analog can be set up. In the case of a piezo-element it is a capacitance with RCL-circuits in parallel, each representing a resonance of the sensor. 
\begin{figure}[ht]
\begin{center}
	\includegraphics[width=6cm]{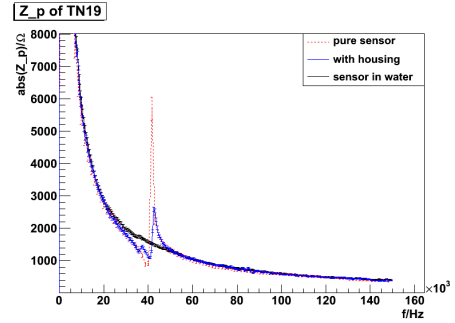}
	\caption{The comparison of the impedance in air and water, showing the reduced resonance in water. "Pure sensor" indicates the status as delivered by Murata. The "Housing" is added in the AAL for the fixing on the positioning system. }
	\label{fig_air_ice_fit}
	\end{center}
\end{figure}

\begin{figure}[ht]
	\centering
 	\includegraphics[width=6cm]{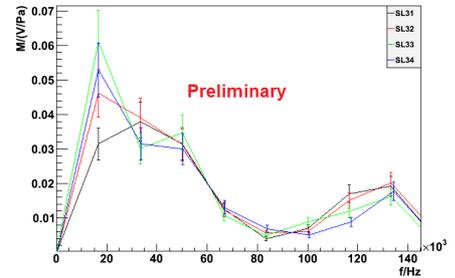}
	\caption{ The sensitivity of the four sensors at level 3 in water.}
	\label{fig_all_rec_water}
\end{figure}

\begin{figure}[ht]
	\centering
	\includegraphics[width=6cm]{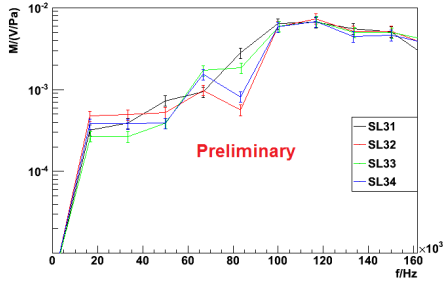}
	\caption{ The sensitivity of the four sensors in level 3 in ice.}
	\label{fig_all_rec_ice}
\end{figure}

In the case of the Murata sensors in water and ice, the representation is reduced to a simple capacitor due to the reduced resonances, see Fig. \ref{fig_air_ice_fit} for water. Results for the sensors in water and ice are shown in Fig. \ref{fig_all_rec_water} and in Fig. \ref{fig_all_rec_ice}.


For the calibration in ice an extra sender is needed, to get a strong input signal at low frequencies. For this reason an underwater loudspeaker is used, installed in the middle at the bottom of the IceTop-tank. This loudspeaker is designed for high emission power at low frequencies (f$<$30$\,$kHz).\\
The systematic errors of the measurements are calculated individually for each sender to $15\%$ and for each sensor to $13\%$. The global error on the calibration measurements is $5\%$ for the senders and $11\%$ for the sensors. The uncertainties are dominated by three components: the error on the angular dependence of the sensitivity and emissivity, the error on the capacitance of the piezo and the error on the reciprocity factor \cite{Thomas}. The last one will be reduced in a  future setup by including senders following a different reciprocity model with less systematic dependencies. The error on the sender impedance can be reduced by a factor of two, by using an upgraded type of sender. The error on the angular dependence can be reduced by measuring a frequency-dependent angular factor for each sensor individually \cite{Benni}. This, especially in ice, is complicated, but will be done for the required angles in future measurements. These systematic errors are dominant for the calibration apart from the in ice calibration of the senders. The low power of the senders in ice leads to statistical uncertainties of about 50$\%$ \cite{Thomas}.
\subsection{Calibration of a SPATS sensor}

The SPATS sensor was developed for the South Pole Acoustic Test Setup \cite{Boeser:2006fx} deployed at South Pole. It consists of three piezo elements pressed against the inner steel housing, a cylinder with a diameter of 10.1$\,$cm. For readout the voltage produced by the piezos is amplified, and sent to a highpass filter (f$>$5kHz) and a lowpass filter(f$<$100$\,$kHz). The SPATS sensor was installed at an ice depth of 45$\,$cm directly in the center of the tank. \\
The sensitivity of the SPATS sensor is calculated using $M = U/p$. The pressure is calculated using the known sensitivity of the AAL sender. It is given by: 
\begin{eqnarray}
 p = \frac{S_t I}{d} = \frac{M_t I}{J d} 
\end{eqnarray}

\begin{figure}[ht]
	\centering
 		\includegraphics[width=6cm]{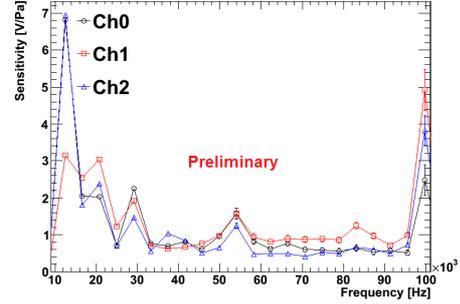}
 		\caption{Sensitivity for the three channels of the SPATS sensor in water.}
	\label{fig_14}
\end{figure}
\begin{figure}[ht]
	\centering
 	\includegraphics[width=6cm]{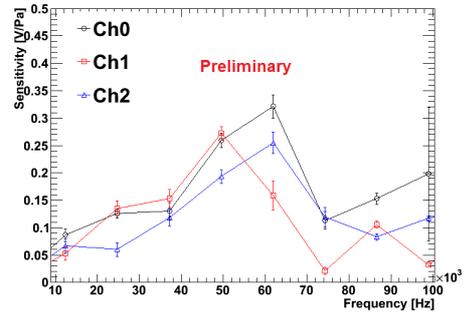}
	\caption{Sensitivity for the three channels of the SPATS sensor in ice.}
	\label{fig_15}
\end{figure}
The sensitivities for one SPATS sensor in water are shown in Fig. \ref{fig_14}. The calibration measurement in water is consistent with the results in \cite{Fischer}. The systematic errors are 23$\%$. They are dominated by the error of the senders sensitivities in Sec. \ref{recalibration}. The peaks in the spectrum at low frequencies arise from resonances of the steel housing and are not related to the intrinsic piezo resonances.\\
We observe a strong reduction of the sensitivity from water to ice for all frequencies $<$ 100kHz, shown in Fig. \ref{fig_15}. This result has not been confirmed in  an independent measurement yet. A problem for the in-ice calibration is the extremely low emissivity of the Murata senders in ice, which results in large systematic uncertainties of 56$\%$ \cite{Larissa}. The loudspeaker could not be used for the measurements reported here, since it is positioned directly beneath the SPATS sensor. It is planned to redo the measurement with improved senders, which have a higher emissivity.

\section{Resonant Sensor}\label{epflSct}
\begin{figure}[ht]
\centering\hskip6mm\includegraphics[width=6.8cm]{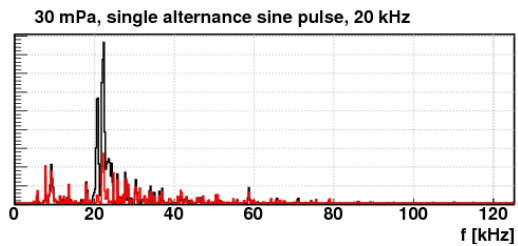}
\caption{Sensor response (power spectrum squared) to a 30 mPa pulse (20 kHz single sine period). Typical noise in red.}
\label{sensorResponse}
\end{figure}

\begin{figure}[ht]
\centering\includegraphics[width=7.5cm]{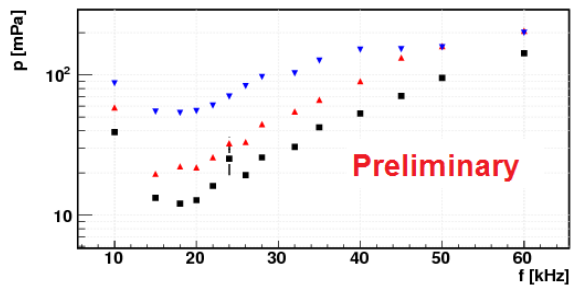}
\caption{Equivalent pressure of the self noise w.r.t. the frequency of a single sine period excitation pulse for three channels.}
\label{sensorSelfNoise}
\end{figure}

A multi-channel acoustic PZT-based sensor encased in a stainless steel housing was recently developed at \'Ecole Polytechnique F\'ed\'erale de Lausanne (EPFL). The overall design is similar to the design of the sensors used by SPATS~\cite{Boeser:2006fx}. Immersed in water, the sensor housing exhibits strong resonant behavior at a fundamental frequency of about 23 kHz. In an ice environment with higher speed of sound however, one expects the resonance to be damped due to a closer impedance match with the surrounding medium. The sensor dimensions were chosen to match well the expected main frequency content of an acoustic pulse generated by an UHE neutrino in ice~\cite{learned:1979,bevan:2007}. This sensor is not likely to perform as well as a sensor with a flat frequency response curve  for the characterization of the exact frequency content of a short transient excitation. On the other hand, the rather sharp frequency response allows to reduce the bandwidth in the PZT amplifiers, thereby decreasing electrical noise.\\
Two versions of the resonant sensor exist, with digital or analog transmission lines. The  digital version has two parallel 8 MHz digital transmission lines which have now been successfully tested over 500 m ethernet cables (each transmission line carries the multiplexed signal of two sensor channels and a 12-bit digitizer samples each channel signal at the rate of 250 MSPS). More details on the sensor layout, components and electronics can be found in \cite{icrc09-ribordy-podgorski}, as well as a discussion of its pointing capabilities (with the sensor encased in an aluminium housing instead).\\
Fig.~\ref{sensorResponse} shows the frequency response of one of the sensor channels to a 30 mPa signal (single sine period, 20 kHz) and for comparison the noise from a single 8.192 ms data acquisition sample in open water. Fig.~\ref{sensorSelfNoise} shows the self noise of the sensor at frequencies between 10 and 60 kHz.

\section{Outlook}
The activities for acoustic neutrino detection at the South Pole continue. Sensors from EPFL and modified SPATS sensors, supplemented with a floating PZT emitter in their center for absolute calibration are currently undergoing intensive testing (including freezing tests and calibration in ice) at the AAL facility. In the context of the continued SPATS project aiming for the absolute acoustic noise level measurement we anticipate the deployment of additional sensors at the South Pole this winter.





\bibliographystyle{elsarticle-num}







\end{document}